\documentclass[12pt]{article}
\RequirePackage{setspace}
\AtBeginDocument{\setlength\abovedisplayskip{3pt}}
\AtBeginDocument{\setlength\belowdisplayskip{3pt}}
\usepackage{amsthm}
\usepackage{amsfonts}
\usepackage{bbm}
\usepackage{graphicx}
\usepackage{sectsty}
\sectionfont{\normalsize}
\usepackage{appendix}
\usepackage{float}
\usepackage{url}
\usepackage{caption}
\usepackage{lineno}
\usepackage[round]{natbib}

\begin{document}
\title{From Quantum Axiomatics to Quantum Conceptuality\footnote{Submitted to a special issue of \emph{Activitas Nervosa Superior: Brain, Mind and Cognition}, dedicated to Henry Stapp in honor of his 90th birthday.}}
\author{Diederik Aerts$^1$, Massimiliano Sassoli de Bianchi$^{1,2}$\\
Sandro Sozzo$^{3}$ and Tomas Veloz$^{1,4,5}$ \vspace{0.5 cm} \\ 
\normalsize\itshape
$^1$ Center Leo Apostel for Interdisciplinary Studies, 
Brussels Free University \\ 
\normalsize\itshape
Krijgskundestraat 33, 1160 Brussels, Belgium \\
\normalsize
E-Mails: \url{diraerts@vub.ac.be}, \url{msassoli@vub.ac.be}
\vspace{0.1 cm} \\ 
\normalsize\itshape
$^2$ Laboratorio di Autoricerca di Base, \\
\normalsize\itshape
Lugano, Switzerland\\
\normalsize
E-Mail: \url{autoricerca@gmail.com}
\vspace{0.1 cm} \\ 
\normalsize\itshape
$^3$ School of Business and IQSCS, University of Leicester \\ 
\normalsize\itshape
University Road, LE1 7RH Leicester, United Kingdom \\
\normalsize
E-Mail: \url{ss831@le.ac.uk} 
\vspace{0.1 cm} \\ 
\normalsize\itshape
$^4$ Instituto de Filosof{\' i}a y Ciencias de la Complejidad IFICC, \\ 
\normalsize\itshape
Los Alerces 3024, \~Nu\~noa, Santiago, Chile
\vspace{0.1 cm} \\
\normalsize\itshape
$^5$ Departamento Ciencias Biol\'ogicas, Facultad Ciencias de la vida\\ 
\normalsize\itshape 
Universidad Andres Bello,  8370146 Santiago, Chile
\\
\normalsize
E-Mail: \url{tveloz@gmail.com}
}
\date{}
\maketitle
\vspace{-0.8 cm}
\begin{abstract}
\noindent Since its inception, many physicists have seen in quantum mechanics the possibility, if not the necessity, of bringing cognitive aspects into the play, which were instead absent, or unnoticed, in the previous classical theories. In this article, we outline the path that led us to support the hypothesis that our physical reality is fundamentally conceptual-like and cognitivistic-like. However, contrary to the `abstract ego hypothesis' introduced by John von Neumann and further explored, in more recent times, by Henry Stapp, our approach does not rely on the measurement problem as expressing a possible `gap in physical causation', which would point to a reality lying beyond the mind-matter distinction. On the contrary, in our approach the measurement problem is considered to be essentially solved, at least for what concerns the origin of quantum probabilities, which we have reasons to believe they would be epistemic. Our conclusion that conceptuality and cognition would be an integral part of all physical processes comes instead from the observation of the striking similarities between the non-spatial behavior of the quantum micro-physical entities and that of the human concepts. This gave birth to a new interpretation of quantum mechanics, called the `conceptualistic interpretation', currently under investigation within our group in Brussels.
\end{abstract}
\\

\section{Introduction}
Following the traces left by pioneers such as Niels Bohr, Werner Heisenberg, John von Neumann and Eugene Wigner, Henry Stapp has been a staunch defender, over the years, of the idea that what quantum theory has revealed to us, among other things, is that reality cannot consists of mindless physical entities (like particles and fields), as the deterministic edifice of classical physics has led us to believe. According to Stapp, the most radical shift brought by quantum physics is the explicit introduction of mind into the basic conceptual structure of the theory. For instance, he saw in the dimension of potentiality a signature of the fact that what reality is made of is more akin to a conceptual/cognitive stuff, formed for instance by `ideas of what might happen', rather than to the persisting substances posited by Newtonian (pre-quantum) theories \citep{Stapp2009,Stapp2011}.

Our group in Brussels was also led in recent years to consider the possibility that `the stuff our reality is made of is fundamentally conceptual'. As for Stapp, we arrived at this view by taking very seriously the quantum formalism, in particular the fact that the two processes that quantum theory contemplates 
-- one deterministic, described by the Schr\"odinger equation, and the other indeterministic, governed by the projection postulate and the Born rule -- are both fundamental. The ontology that emerges from our analysis is however quite different from that presupposed by Stapp -- although not necessarily incompatible with it -- and it is the purpose of this article to trace the trajectory of the ideas that led us to consider this conceptualistic paradigm shift.

\section{Quantum axiomatics}
It began with \citet{BirkhoffvonNeumann1936} instigating a domain of research that was able to achieve a number of interesting results in the seventies and eighties of the last century. At that time, pioneers like \citet{Mackey1963}, \citet{Jauch1968}, \citet{Piron1976}, \citet{FoulisRandall1978} and \citet{Ludwig1983}, were all dedicated to the task of building an axiomatic foundation for quantum theory starting from as much as possible operationally founded axioms, also allowing in this way for structurally more general quantum-like theories, in case part of the proposed axioms were withheld. One of us, student of Constantin Piron, was also thoroughly engaged in this research on quantum axiomatics in the foregoing century \citep{Aerts1982a, Aerts1986, Aerts1999a, AertsDurt1994, Aertsetal1997a, Aertsetal1999b}, so much so that the Geneva school of quantum mechanics became also known -- at least among the insiders -- as the `Geneva-Brussels school'.

As we mentioned, an important aspect of these axiomatic approaches is the possibility to frame standard quantum mechanics within more general mathematical structures, able to jointly describe quantum and classical systems, understanding them as limit situations of more general `in between quantum and classical' intermediary systems. This allowed in particular to appreciate a key point: physical entities have properties that can be classical, quantum or intermediate, and that this does not depend on the fact that they would be microscopic or macroscopic, or immersed in a cold or hot environment. In other words, and more precisely, the realization was that what we came to call ``quantum'' is first of all a form of organization, which can be present in certain experimental contexts and revealed by performing suitably designed experiments. 

\section{Quantum machines}
One of the consequences of this understanding was that, being the appearance of the quantum laws not limited to the micro-level and to the very low temperature regimes, it was possible to conceive a number of macroscopic physical systems with a quantum behavior, due to the specific way measurements were defined with respect to them. In other words, these were physical systems of a classical nature which could acquire a `quantum-like nature' when some specific measurements, perfectly defined in operational terms, were replacing the classical ones. It could indeed be proved that these macroscopic quantum-like physical systems entailed non-Kolmogorovian probability models, which made them a very valuable study object for the investigation of possible origins of non-Kolmogorovity and quantumness \citep{Aerts1986,Aertsetal1997b,Aerts1998,Aerts1999b,SassolideBianchi2013a, SassolideBianchi2013b}. 

More specifically, these ``quantum machines'' were analyzed in great depth with the aim to shed light on the origin of quantum indeterminism. It was well known, thanks to the existing no-go theorems restricting the permissible hidden-variable models \citep{Neumann1932,Bell1966,Gleason1957,Jauch1963,Kochen1967,Gudder1970}, that quantum probabilities could not reflect a situation of lack of knowledge about `better defined states' of a quantum entity. As a consequence, the majority of physicists was led to believe that quantum probabilities are `ontic', and not `epistemic', i.e., not explainable as a lack of knowledge about an objective deeper reality. But the quantum machines unveiled a completely different scenario that opened to a new possibility.

Indeed, because of their macroscopic nature, and the explicit way in which measurements were defined, everything was perfectly visible about how these machines operate. What they showed is that the fluctuations at the origin of the quantum probabilities were not to be attributed to the states of the measured systems, which were perfectly well defined prior to the measurements, but to their interactions with the measuring apparatuses. The no-go theorems, however, assumed that the hidden-variables were to be assigned to the states of the entities under consideration, so they did not apply anymore if they were instead assigned to the measuring interactions \citep{Aerts1984}. More interesting, the insights gathered from these macroscopic quantum machines allowed for the identification of a theoretical approach to derive the Born rule in a non-circular way, first for two-dimensional systems \citep{Aerts1986} and then, in more recent times, for systems of arbitrary dimension, in what has now been called the `extended Bloch representation of quantum mechanics' \citep{AertsSassolideBianchi2014,AertsSassoli2016,AertsSassoli2018}. 

\section{Observations and creations}
The important aspect that was revealed by these studies is that in the description of a physical entity one needs to distinguish between two typologies of lack of knowledge. One is about an incomplete knowledge of the actual state of the entity, and the other is about an incomplete knowledge of the interactions arising between the entity and its experimental context, particularly the measuring apparatus. Classical probabilities, obeying Kolmogorov's axioms, formalize the former ignorance, whereas quantum probabilities, not obeying Kolmogorov's axioms, formalize the latter, and more precisely, they correspond to the limit situation where there is a complete knowledge of the state of the entity but a maximum lack of knowledge about the interaction that is actualized with the measurement apparatus. And as mentioned above, in between these two extremes one can find intermediate situations, giving rise to non-classical (non-Kolmogorovian) and non-quantum (non-Hilbertian) probability models. 

In other words, as the very formalism of quantum mechanics already indicates, and the quantum machines confirmed, quantum measurements are not just observations without effects: they usually provoke real changes in the state of the entity and are intrinsically invasive, as they `create properties' that were not actual for the measured entity prior to the execution of the measurement' This `actualization of potential properties' works  like a weighted symmetry breaking process, during which one of the available hidden measurement-interactions -- each one being associated with a deterministic process -- is actualized \citep{AertsSassolideBianchi2017}. It is really `the actual that breaks the symmetry of the potential'. Interestingly, the reason why quantum measurements would be genuinely indeterministic is not because a hidden-measurements understanding of them would be incompatible with the view of determinism applying to the whole of reality, but because if we would remove the fluctuations -- and therefore the randomness -- which is built in certain of our observational processes, acquiring on them a better control, we would also irredeemably alter these same processes, so that in the end they would not correspond anymore to the test/observation of the same properties \citep{SassolideBianchi2015}.

\section{Non-spatiality}
At this point of our story, one may wonder how the idea that our reality would be fundamentally conceptual could emerge. Indeed, our narrative goes precisely in the direction of demystifying the celebrated measurement problem, showing that no psychophysical argument {\`a} la von Neumann/Stapp would be required to explain the reduction of the state vector during a measurement. To present an analogy, one would not dare introducing an `abstract ego {\` a} la von Neumann' to explain why, when we push on a cylindrical stick vertically planted in the ground, the process will actualize a bending direction -- breaking the initial rotational symmetry -- which was only potential prior to the pushing experiment. Of course, we always need an experimenter choosing which measurement to perform and when, and certainly this act of choice, if truly free, might well go beyond the possibilities of a scientific description.\footnote{Regarding this free choice aspect, \citet{Stapp2009} writes: ``This choice is sometimes called the `Heisenberg choice', because Heisenberg emphasized strongly its crucial role in quantum dynamics. At the pragmatic level it is a `free choice', because it is controlled, at least at the practical level, by the conscious intentions of the experimenter/participant: neither the Copenhagen nor von Neumann formulations specify the causal origins of this choice, apart from the conscious intentions of the human agent.} But this being true or not, it has little to do with quantum mechanics per se, considering that, as we mentioned already, there is no incompatibility between the latter and determinism, as all interactions between physical entities and measuring apparatuses can be assumed to be deterministic -- which of course does not mean they necessarily are.

Now, if we accept the above-mentioned `hidden-measurements' explanation for the indeterminism in quantum measurements, we should not forget that this also applies to position measurements. Also during a position measurement the outcome will be determined by the interactions taking place between the measuring apparatus -- a detector in a laboratory -- and the quantum micro-entity, with the latter being usually, before the measurement, in a superposition state `not corresponding to an actual position', which is then changed by the measurement into an eigenstate expressing an `actual position'. This means that the quantum entity before the position measurement `had no position', or `was non-spatial', if we call `space' the set of all possible positions. 

To understand the paradigmatic step we made towards considering the stuff reality is made of to be conceptual -- its specific nature being guided by what we learned from the domain of research called `quantum cognition', as we will explain later in the text, -- we cannot emphasize enough how important it was to introduce the notion of `non-spatiality' for the default state of a quantum micro-entity, and to attach an objective reality to it, i.e., a quantum entity in such a state is simply `not inside the three-dimensional realm which we call space'. Hence, a quantum entity only becomes `spatial', i.e., manifests inside space, when forced by a position measurement to enter into a `detected state'. So, our demystification of the quantum measurements brings with it a new mysterious element about our physical reality: the non-spatial nature of quantum states in general and, as a consequence, `space being too limited to be a theater for the overall reality of the quantum entities' \citep{Aerts1999b,AertsSassolideBianchi2017c}.

We have to mention two additional points that played an important role in the genesis of our `conceptuality interpretation' of quantum theory. The first one is of an experimental nature, while the second one is theoretical, and both contributed in an important way to the bold move of considering quantum entities to be non-spatial. Before the hidden-measurements explanation was formulated, the two experimental groups guided by Helmut Rauch and Samuel Werner started using a Si-crystal interferometer to observe the $4\pi$-periodicity of the spinor wave functions, when rotated by a static magnetic field \citep{Rauchetal1975,Werneretal1975}. These experiments showed that it was possible to create interference effects by acting only on one of the beams inside the interferometer, even when a single neutron at a time was present in it. When attentively analyzed, these results force us to go beyond the na{\"\i}f wave-particle duality, which is a spatial duality, showing that neutrons can be prepared in states such that they can be influenced by local apparatuses (like magnetic fields) acting on separated regions of space, without ever manifesting in between these regions and still remaining whole local entities when they are brought into a `detected state' 
\citep{Aerts1999b,SassolideBianchi2017}. This means that the quantum micro-entities like neutrons cannot be considered to be customarily present in space, although they always maintain a relation with it, by remaining available to be drawn into it by a measuring apparatus, when certain conditions are met.

\section{Separate entities}
The second point that also played an important role in considering quantum entities to be primarily non-spatial is the theoretical study of the simple situation of `two separate entities', which was studied within the axiomatic operational-realistic approach developed in the Geneva school on the foundations of quantum theory. The mathematical structure needed to model this situation could be worked out explicitly and the surprising discovery was that two of the well-known axioms of standard quantum theory -- in its lattice structure formulation -- called the `weak modularity' and `covering law' axioms, were never satisfied for two separate quantum entities. Since this result was proven in a rigorous way, a direct consequence of it was the veracity of the following statement: `standard quantum theory is incapable of modeling two separate quantum entities'. 

This investigation on the axiomatics of separate quantum entities took place in the same period when the first experiments on the EPR-paradox saw the light, and there was a clear connection between these two situations, as analyzed in \citet{Aerts1982a,Aerts1984} and \cite{SassolideBianchi2018}. More precisely, it could be shown that the `incompleteness' evidenced by the EPR-reasoning was not an incompleteness in the sense of the hidden-variables to be associated with the quantum states, but an incompleteness in the sense of an incapacity of modeling separate quantum entities. This result was not determinative in itself for considering quantum micro-entities to be non-spatial, but it proved important in relation to the other results. Indeed, the hidden-measurements interpretation applied to position measurements was a strong theoretical indication of the non-spatial nature of quantum states. Rauch and Werner experiments with neutron interferometry were then able to directly illustrate some important aspects of such remarkable non-spatiality. And the results on separate systems and quantum axiomatics provided an indication that what is today commonly referred to as `non-locality' is not a feature specifically related to `distant regions of space', but a much more general and structurally deep aspect of quantum entities. 

In fact, this can already be felt at the level of the standard quantum formalism, where entanglement comes about as a direct consequence of the superposition principle also applying to systems formed by quantum sub-entities. Let us mention here that one of the two `failing axioms' in the study of separate quantum entities, `the covering law', is the equivalent of the superposition principle. This proves on a deep axiomatic level that it is indeed the superposition principle that is at the root of entanglement, and the quantum phenomena revealed in experiments on `single quantum entities', like those of Rauch and Werner,  are actually the same as those revealed in EPR-like experiments, involving bipartite systems. Hence,  putting together the hidden-measurements explanation, the neutron interferometry experiments, and the axiomatic study of the situation of separate quantum entities, this made us make the step to consider non-locality as a specific manifestation of non-spatiality and to consider the latter as a fundamental property of quantum entities.

\section{Quantum cognition}
We reach the next crucial part of our story, regarding the emergent domain today known as `quantum cognition', which is about using the quantum formalism and its ensuing quantum probability to model the experimental data obtained in psychological measurements, particularly those that could not be modeled in the past by means of classical logical and probabilistic structures. This domain was initiated in the nineties of the last century by one of us and his group of collaborators in Brussels, like Sven Aerts, Liane Gabora, Jan Broekaert, Sonja Smets and Marek Czachor \citep{aertsaerts1995,aertsbroekaertsmets1999,gaboraaerts2002,aertsczachor2004,AertsGabora2005a,AertsGabora2005b}, as well as by \citet{khrennikov1999} and \citet{altmanspacher2002}. All these researchers were influenced by the previously mentioned advancements in the foundations of quantum theory and so-called `quantum logic', which brought in particular the Brussels group to introduce a generalized quantum theory, called the `SCoP (State Context Property) formalism', particularly suited for the modeling not only of physical entities, but also of abstract entities, like human concepts \citep{AertsGabora2005a,AertsGabora2005b}.

The SCoP formalism was much more general than quantum mechanics, and in the beginning it was believed that the standard quantum formalism was too specific to describe the full structural complexity of the contextual effects produced by psychological experiments, which in fact was recently demonstrated to be the case \citep{Aertsetal2016,AertsSassolideBianchi2017b,Aertsetal2018b}. But despite these `beyond-quantum anomalies', and certainly beyond the initial expectations, pure Hilbertian structures, and the associated Born rule for calculating the probabilities of the outcomes, appeared to be extremely effective in the modeling of the main quantum effects identified in these domains a priori very different from the micro-world \citep{AertsSozzo2011,AertsSozzo2014,AertsSozzoVeloz2015}. So, why the quantum formalism turned out to be so successful in the modeling of cognitive phenomena? 

In a nutshell, quoting from \cite{AertsSassoli2018b}, because: ``human minds deal essentially with \emph{concepts}, which are highly contextual entities, and that quantum mechanics is a theory that has precisely been designed to deal with \emph{contextuality}. Also, in the same way a quantum entity can be in different states, a concept can also be understood as a \emph{meaning entity} whose states depend on the (semantic) context in which it is immersed. Sometimes contexts will influence concepts (i.e., change their meanings, and therefore their states) in a deterministic way, other times in a perfectly indeterministic way, for instance when a mind is put in the situation of answering a question without having already a preprogrammed response, so that the latter must be created at the moment, similarly to how a potential outcome is actualized during a quantum measurement. Furthermore, concepts can form \emph{connections through meaning}, which in turn can produce significant correlations when these meaning-connections are tested/actualized in specific experimental situations. Thus, in the conceptual (human) realm, similarly to physics, one can design experimental situation where Bell's inequalities can be violated, showing that concepts can entangle in similar ways as quantum entities can do \citep{AertsSozzo2011,AertsSozzo2014}. Then, there is also the fact that when concepts are combined, new meanings can easily emerge in ways that cannot be described by considering the classical (Aristotelian) view that concepts would be just like containers of exemplars. These non-compositional emergence effects produced by conceptual combinations, when analyzed in statistical terms by performing experiments with large groups of subjects, can again be shown to be like the quantum mechanical interference effects, resulting from the superposition principle \citep{AertsSozzoVeloz2015}.''

\section{Quantum conceptuality}
The above passage does not describe all the correspondences evidenced so far between the behavior of quantum physical entities and that of human concepts, and we refer the interested reader to the today available vast literature on quantum cognition, to fully appreciate how much `quantumness' has been discovered, at the structural level, in human cognitive processes \citep{Khrennikov2010,BusemeyerBruza2012,HavenKhrennikov2013,Wendt2015}. These remarkable correspondences did not go unnoticed by one of us, who formulated a decade ago a thought-provoking question: Considering that quantum mechanics has been applied with such success to the modeling of human concepts and their interaction with human minds, could this mean that the micro-physical entities would possess a conceptual nature? In other words, could it be that `quantumness' and `conceptuality' are just two terms referring to a same reality, or nature, which manifests at different organizational levels within our complex reality? This bold hypothesis was explored in a number of papers, in what has now been called the `conceptuality interpretation of quantum mechanics' \citep{Aerts2009,Aerts2010a,Aerts2010b,Aerts2013,Aerts2014,Aertsetal2017}, according to which the nature of quantum entities would be conceptual, in the sense that they would interact among them and with the measuring apparatuses (entities made of ordinary matter) in an analogous way as human concepts combine with each other in our linguistic constructions and interact with human minds (memory structures sensitive to the meaning carried by concepts). 

Clearly, it was not the purpose of this article to present the necessary technical arguments in favor of the conceptuality interpretation of quantum mechanics, -- to be found in the above-mentioned articles, -- which is still under intense investigation, as we believe we have only touched on its full potential as far as its explanatory and clarifying power is concerned. Our intent here was to present the concatenation of ideas that led us to consider a conceptualistic paradigm shift as a serious possibility. The consequence of it, in case it would turn out to be correct, is that our view of inert matter must change profoundly, as even though deprived of sensorineural apparatuses, it would be however able to perceive/create with the surrounding material environment similarly to how human minds/brains are able to perceive/create with their conceptual environments. The resulting worldview is one that we have called `pancognitivism', stating that everything within reality, being fundamentally conceptual, would participate in cognition, with human cognition being just an example of it, expressed at a very specific organizational level \citep{AertsSozzo2015,AertsSassoli2018b}. 

This pancognitivist view, however, is not a trivial one and the mistake not to commit is to use it to promote a radical anthropomorphic view of reality. Indeed, in the same way the notion of wave describes both the beingness of an electromagnetic wave and of a sound wave, which otherwise remain very different entities, it is the notion giving rise to the beingness of a quantum entity and of a human concept that are here assumed to be the same, whilst a quantum entity, like a photon or an electron, clearly remains very different from a human concept, and pieces of ordinary matter remain very different from a human mind/brain systems. What in physics we indicate as non-locality, and should in our view  more properly 
be considered non-spatiality, would then just be an expression of the fact that quantum entities, as conceptual entities, can be in more or less `abstract states', with the most concrete states corresponding to the `spatial (detected) states'. Heisenberg's uncertainty principle then becomes an ontological statement expressing the necessary tradeoff between concreteness and abstractness, resulting from the fact that, at the ontological level, quantum entities would be conceptual entities that cannot be simultaneously in a state of maximal abstractness and maximal concreteness \citep{Aertsetal2017}.

\section{Conclusion}
To conclude, we observe that Henry Stapp considered the quantum reduction events to be the physical manifestation of (the final part of) psychophysical processes, where there would be a correlation between the psychological aspects associated with the experimenters' free choices and von Neumann's process 1 physical processes. In other words, for Stapp the physical and psychological aspects of our reality would be intimately bound together in the quantum events, so much so that orthodox quantum mechanics, while remaining Cartesian dualistic at the pragmatic and operational level, would be mentalistic at the ontological level. Different from our pancognitivistic view, which results from an observation of the similarity of behaviors between quantum and human conceptual entities, and the fact that micro-entities appear to be non-spatial (abstract) entities, Stapp's version of panpsychism results from an examination of the causal structure of the theory. These two ontological interpretations of quantum mechanics are therefore very different; nonetheless, they reach non-mutually exclusive conclusions regarding the nature of the stuff our world is made of.

\end{document}